\documentclass[useAMS,usenatbib,fleqn]{mn2e}
\usepackage{graphicx,amsmath,amssymb,times}
\newcommand{\beq}{\begin{equation}}
\newcommand{\eeq}{\end{equation}}
\title[Possible Proton Synchrotron Origin of ..... ]
  {Possible Proton Synchrotron Origin of X-Ray \& Gamma Ray Emission in Large Scale Jet of 3C 273}
\author[Esha Kundu and Nayantara Gupta]
  {Esha Kundu$^1$\thanks{esha.kundu@gmail.com} 
  and Nayantara Gupta$^2$\thanks{nayan@rri.res.in}\\   
  $^1$Tata Institute of Fundamental Research, Homi Bhabha Road, Colaba, Mumbai 400005, India\\
  $^2$Raman Research Institute, Sadashiva Nagar, Bangalore 560080, India }
\date{\today}

\pagerange{{00}--{00}} \pubyear{0000}

\def\LaTeX{L\kern-.36em\raise.3ex\hbox{a}\kern-.15em
  T\kern-.1667em\lower.7ex\hbox{E}\kern-.125emX}

\begin{document}

\label{firstpage}

\maketitle

\begin{abstract}
The  large scale jet of quasar 3C 273 has been observed in radio to $\gamma$ ray frequencies. Earlier the X-ray emission from knot A of this jet has been explained with inverse Compton scattering of the cosmic microwave background radiations by the shock accelerated relativistic electrons in the jet. 
More recently it has been shown that this mechanism overproduces the gamma ray flux at GeV energy and violates the observational results from Fermi LAT.
We have considered the synchrotron emission from a broken power law spectrum of accelerated protons in the jet to explain the observed X-ray to $\gamma$ ray flux from knot A. The two scenarios discussed in our work are
 (i) magnetic field is high, synchrotron energy loss time of the protons is shorter than their escape time from the knot region and the age of the jet (ii) their escape time is shorter than their synchrotron energy loss time and the age of the jet. These scenarios can explain the observed photon spectrum well for moderate values of Doppler factor. The required jet luminosity is high $\sim 10^{46}$ erg/sec in the first scenario and moderate $\sim 10^{45}$ erg/sec in the second, which makes the second scenario  more favorable.   
\end{abstract}

\begin{keywords}
 quasars:general; X-rays: galaxies; gamma-rays:galaxies
\end{keywords}

\section{Introduction}
The large scale jet of the quasar 3C 273 has been observed in
radio to $\gamma$ ray frequencies by different ground based and space based 
detectors. The optical image taken by HST \citep{bah} showed the presence of bright knots in the kpc jet of 3C 273. The first knot, called kont A (knot nomenclature according to \citet{jester}), is the most luminous one in X-rays.

The emission in the radio to optical frequencies from knot A is consistent with synchrotron radiation of relativistic electrons in the jet \citep{sam1} whereas the X-ray emission has been modelled by the inverse Compton 
scattering of the cosmic microwave background photons (IC/CMB) by the relativistic electrons \citep{sam2}. It has been shown that the synchrotron self Compton mechanism can not explain the observed X-ray flux \citep{sam1}.    \\
The IC/CMB model requires the jet to be highly collimated upto the location of the X-ray knot pointing close to our line of sight or the Lorentz factor has to be high. The small angle subtended by the jet to our line of sight may require the length of the jet to be Mpc in some cases. The electron spectrum in the jet has to extend down to 1-10 MeV.
Also sometimes this model requires huge jet kinetic power exceeding the Eddington power of the source \citep{dermer,uchi} to explain the observed X-ray fluxes. X-ray emission could also be possible from synchrotron emission from an additional shock accelerated electron population \citep{hard, jester, uchi}. 
Although the synchrotron emission mechanism is far more efficient than IC/CMB
without requiring high Lorentz factors, extreme jet lengths, high jet kinetic powers to explain the observed X-ray fluxes, it is not yet known which physical mechanism might produce the second population of electrons or additional EED.  
\citet{meyer} have analyzed more than four years of Fermi LAT $\gamma$ ray data from knot A and ruled out the IC/CMB process as a possible mechanism of X-ray radiation as IC/CMB overprdouces the GeV $\gamma$ ray flux.\\
The observed energy spectrum of radio, optical and X-ray photons at 1 keV has an overall spectral index 0.75 which deviates to 0.6 for the local X-ray photons according to Marshall et al. 2001. The synchrotron power per unit frequency 
having a spectral index 0.75 originates from an shock accelerated electron spectrum $\frac{dN_e}{dE_e}\propto E_e^{-1.5}$, which steepens to $E_e^{-2.5}$ due to synchrotron losses. The high energy electrons are expected to lose energy very fast due to synchrotron cooling before moving far from their production site. 
Due to this reason unless the electron acceleration is taking place throughout the entire knot or there are many compact acceleration sites inside the knot it is hard to explain the X-ray emission by synchrotron radiations of high energy electrons. 
Aharonian (2002) proposed synchrotron emission by accelerated protons to be the origin of the radio to X-ray or only the X-ray emission observed from knot A of 3C 273 to overcome the problem of fast cooling of the high energy electrons. The age of the jet ($3\times 10^7$ years), the escape time of the accelerated protons after diffusion and the time for their synchrotron energy loss are the crucial parameters in determining the spectral index of the proton spectrum in this model. \\
In this paper we revisit the proton synchrotron model to identify the possible  common origin of the X-ray and $\gamma$-ray emission from knot A of 3C 273. \\ 

%There can be two different zones present inside the knot which are characterized by different particle population, ambient magnetic 
%field etc. The angle of view of the two zones may be different with respect to an observer on earth. 
AGN jets have been speculated earlier to be one of the potential sites of 
accelerating protons up to $10^{20}$eV \citep{hillas,ces,rachen, henri}. 
\citet{rachen} showed that with size of about 1 kpc and ambient magnetic field of 0.5 mG the jet terminal shock can accelerate proton upto $10^{20}$eV in the hotspot of FRII radio galaxy. In this case the jet can be moderatly relativistic. 
Proton accleration to extremely high energy can also take place at the jet shear bounday layer as demonstrated by \citet{ostro}.  
Protons can be accelerated to $10^{19}$eV \citep{murase} by Fermi acceleration mechanism in the inner jets of radio loud AGN. 
\citet{ebi} has recently proposed that the plasma wakefield, formed by intense electromagnetic field, can accelerate protons/nuclei beyond $10^{21}$eV in the AGN jet over an extended region. 
In this case the particles are accelerated by the Lorentz invariant pondermotive force.
The proton synchrotron models discussed in our work requires the maximum energy of the protons to be close to $10^{20}$ eV.

\section{Proton Synchrotron Spectra}
\subsection{Radio to X-ray Emission}
\par
The X-ray emission from knot A\footnote{\citet{aha} called it knot A1 according to the nomenclature used in \citet{mar}. Note that knot A and knot A1 define the same region of the jet. So hereafter we call it knot A.}
of the large scale jet of 3C 273 has been explained earlier \citep{aha} with synchrotron emission from a relativistic proton population. In this work the author has considered three different scenarios to explain the observed spectral energy distribution in radio to X-ray frequencies. 
The first model (Model I) uses a broken power law spectrum of protons with a spectral index of 2.4 and 3.4 below and above the break energy at $E_b = 3 \times 10^{17}$eV respectively. 
If a proton of energy $E_p$ is trapped in a region of homogeneous magnetic field B and size R then the escape time scale from that region is given by Eq.(2) of \citet{aha} 
\beq
t_{esc} \simeq 4.2 \times 10^{5} \eta^{-1} B_{mG} R_{kpc}^2 E_{19}^{-1} yr 
\label{t_esc}
\eeq
where $\eta$ is known as gyrofactor. In Bohm diffusion limit $\eta = 1$ and in other cases the value of $\eta$ is more than one. 
Here $E_{19} = E_p~ eV/10^{19}$, $B_{mG} = \frac{B}{10^{3}}$G and $R_{kpc} = \frac{R ~cm}{3.08 \times 10^{21}}$. Observational evidences suggest that the size of the knot should be $\leq$ 1kpc. The age of the jet is nearly $3 \times 10^{7}$yr.
For this model the author has assumed that the magnetic field $B = 5$mG and the size of the emission region $R = 1$kpc. In Bohm limit Eq.(\ref{t_esc}) gives $t_{esc} = 2.1 \times 10^{6}$yr for $E_p = 10^{19}$eV. 
The synchrotron cooling time is calculated according to Eq.(1) of \citet{aha0} 
\beq
t_{sync} \simeq  1.4 \times 10^{7} B_{mG}^{-2} E_{19}^{-1} yr
\label{t_sync}
\eeq
For maximum proton energy of $E_{max} = 10^{19}$eV the synchrotron energy loss time $t_{sync} = 5.6 \times 10^5$yr. 
For protons of energy $E_p = 3 \times 10^{17}$eV, $t_{esc}$ and $t_{sync}$ are 
 ($t_{esc} = 7 \times 10^{7}$yr and $t_{sync} = 2 \times 10^7$yr) comparable to the age of the jet.
Above this energy the synchrotron loss dominates over escape loss which makes the injected proton spectrum steeper by $E_p^{-1}$. 
In this model the 
%total energy deposited in protons is $W_{j,p} = 1.14 \times 10^{62}$erg and in magnetic field it is $W_{j,p} = 1.3 \times 10^{59}$erg
% during the last $3 \times 10^7$yr. Therefore 
 required luminosity in cosmic ray protons is $L_{j,p} = 1.2 \times 10^{47}$erg/sec and in the magnetic field it is $L_{j,B} = 1.3 \times 10^{44}$erg/sec for a jet lifetime of $3\times 10^7$yr. This inequality of jet luminosities in protons and the magnetic field implies equipartition of energy does not hold in this case. It is noted that the contribution of low energy protons to the jet luminosity is huge although the protons in the low energy regime are incapable of producing the observed photon fluxes in the radio to X-ray frequencies. 

\par
To reduce the energy budget \citet{aha} has considered another proton spectrum
 with a low energy cut off at $E_{brk} = 10^{13}$eV. It has also been 
 suggested that the energy requirement can be reduced by assuming (Model II) an energy dependent escape time scale which is 
\beq
t_{esc} = \frac{1.4 \times 10^7} {(E/10^{14}eV)^{0.5}} yr
\label{t_esc2}
\eeq
In this model the spectral index of the shock accelerated proton spectrum is 2 
below the break energy ($E_{brk}$). At the break energy $t_{esc} = 4.4 \times 10^7$ yr, using Eq.(\ref{t_esc2}), and $t_{sync} = 1.4 \times 10^{10}$yr.
In this case beyond the break energy the escape loss dominates over the synchrotron loss. Thus the injected proton spectrum $E_p^{-2}$ becomes $E_p^{-2.5}$ above the break energy due to energy dependent losses due to escape.

The upper cut-off energy of the protons in this model is $E_{cut}=10^{18}$eV. 
At the cut-off energy the rate of escape of the protons is ten times higher than their synchrotron cooling rate ($t_{esc} = 1.4 \times 10^{5}$yr, $t_{sync} = 1.4 \times 10^{6}$yr ). %These time scales for different proton energies are tabulated in table \ref{tab0}. 
The value of the ambient magnetic field is assumed to be 10mG in this model. Accelerated protons in the energy window of $10^{13}-10^{18}$eV are mainly responsible for the synchrotron emission in radio to X-ray regime.
For the time period of $1.4 \times 10^{7}$yr the cosmic ray energy budget required for model II is $W_{j,p} = 4.9 \times 10^{60}$erg which results in 
cosmic ray luminosity of $L_{j,p} = 1.1 \times 10^{46}$erg/sec. 
The total energy in magnetic field is $W_{j,B} = 4.9 \times 10^{59}$erg corresponding to a jet luminosity in magnetic field $L_{j,B} = 1.1 \times 10^{45}$erg/sec.
 Thus in model II the jet luminosity in magnetic field is one tenth of that in 
cosmic ray protons.

\par
In comparison to model I the required luminosity in Model II is less but it is still large.
To further reduce the energy requirement \citet{aha} proposed another model (Model III) where the cosmic ray proton spectrum is a power law with a spectral index 2 and an exponential cut off at $E_{cut} = 10^{18}$eV. The magnetic field inside knot A is assumed to be $B = 3$mG and the size of the emission region is $R = 2$kpc.
In this model $t_{sync}$ and $t_{esc}$ are larger than the age of the jet 
for all protons of energy below $E_{cut}$. At $E_{cut}$ the two time scales $t_{sync} = 1.6 \times 10^7$yr and $t_{esc} = 5.1 \times 10^{7}$yr (using Eq.(\ref{t_esc})) are comparable to the age of the jet.  
Due to this reason the spectral index of the shock accelerated proton spectrum 
is not affected by escape or synchrotron losses upto $E_{cut}$ and remains 2 . 
In model III the luminosities in cosmic ray protons and in magnetic field are $L_{j,p} = 10^{45}$erg/sec and $L_{j,B} = 3.7 \times 10^{44}$erg/sec respectively assuming a jet age of $3 \times 10^{7}$yr. We note that 
in this model the synchrotron emission by the relativistic proton population is only capable of producing the observed X-ray flux. The low energy radiation 
is the synchrotron emission from the shock accelerated electrons. 
Figure 3 of \citet{aha} shows the synchrotron spectra produced in model I, II and III discussed above. The different time scales for the 
three models are tabulated in table \ref{tab0} for different proton energies. 

\begin{table}
 \caption{$t_{sync}$ and $t_{esc}$ values in yr for three models of \citet{aha}.
 *using Eq.(\ref{t_esc}). 
 **using Eq.(\ref{t_esc2}).
 }
 \label{tab0}
 \begin{tabular}{@{}lccccc}
  \hline
  Model & $R_{kpc}$ & $B_{mG}$ & $E_{p}$(eV)& $t_{sync}$(yr) & $t_{esc}$(yr) \\
  \hline
  I & 1 & 5 & $3 \times 10^{17}$ & $2 \times 10^7$ & $7 \times 10^{7}$ $^{*}$   \\
  I & 1 & 5 & $10^{19}$ & $5.6 \times 10^5$ & $2.1 \times 10^{6}$ $^{*}$  \\
  \hline
  II & 1 & 10 & $10^{13}$ & $1.4 \times 10^{10}$  & $4.4 \times 10^{7}$ $^{**}$ \\
  II & 1 & 10 & $10^{18}$ & $1.4 \times 10^{6}$  & $1.4 \times 10^{5}$ $^{**}$ \\
  \hline
  III & 2 & 3 & $10^{18}$ & $1.6 \times 10^7$ & $5.04 \times 10^{7}$ $^{*}$ \\
  \hline
 \end{tabular}
\end{table}

\subsection{X-ray \& Gamma ray Spectra} 
\begin{figure}
\centering
 \includegraphics[width=9cm,origin=c]{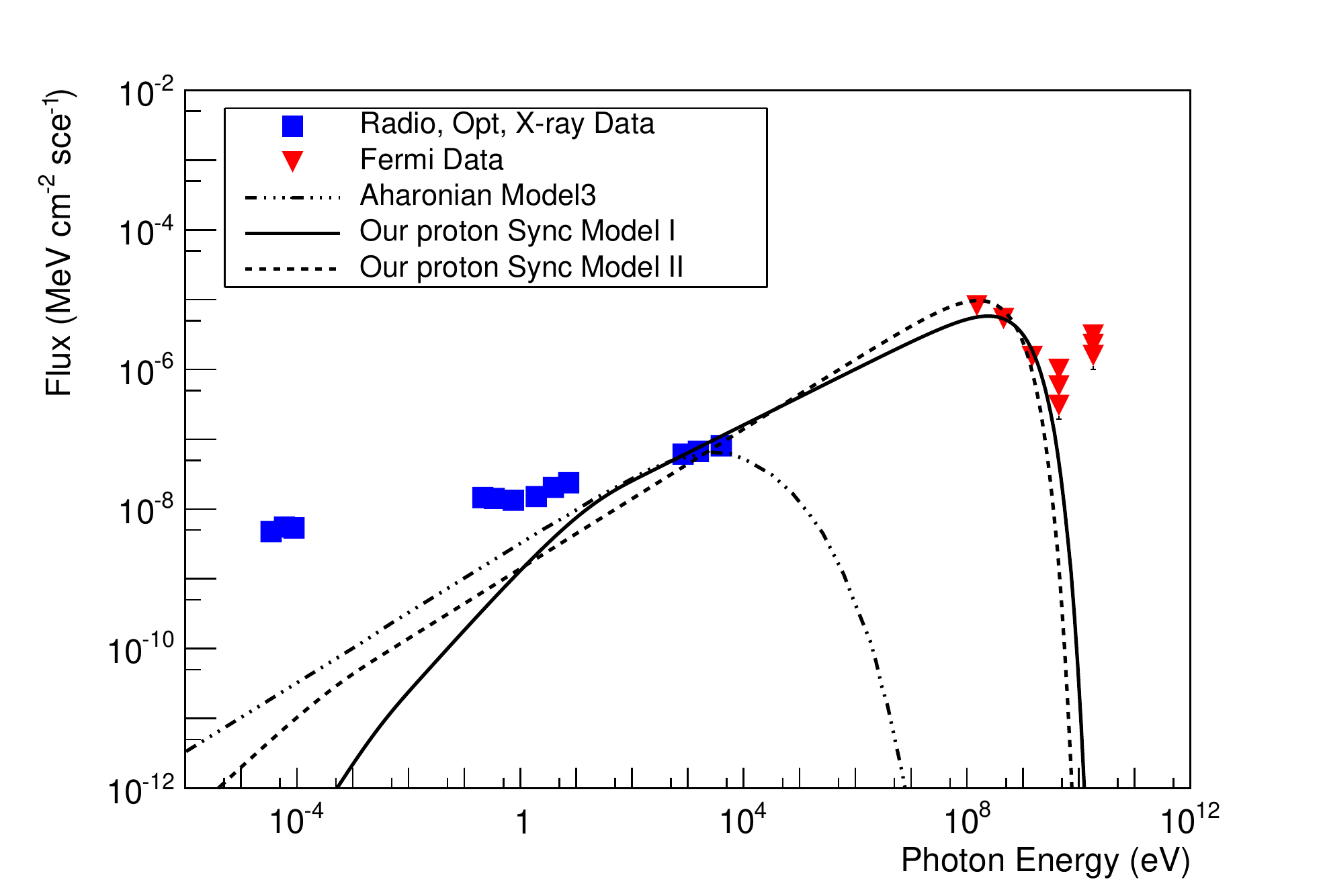}
 \caption{Spectral energy distribution of knot A of 3C 273 shown, radio to X-ray data (\citet{uchi}, \citet{jester0}), gamma ray data and upper limits (\citet{meyer}). Solid line: synchrotron radiation from broken power law proton spectrum corresponding to our model I, dashed line: our model II, dot-dot-dashed line:
 proton synchrotron spectrum corresponding to model III of \citet{aha}. 
Our work uses the code from \citet{kraw}. 
}
 \label{fig1}
\end{figure}

\citet{meyer} have provided the photon fluxes in the three energy windows ($100-300$MeV, $300-1000$MeV \& $1-3$GeV) after analysing the Fermi LAT data from knot A of 3C 273.
For the highest two energy bins, $3-10$GeV \& $10-100$GeV, only upper limits are available due to poor statistics. 
The observed flux and the upper limits are shown in Fig.\ref{fig1} with triangle symbols. This observed flux can not be explained by extrapolating the IC/CMB curve to gamma ray energy. 
We propose two models where proton synchrotron emission by extremely relativistic protons can explain the observed x-ray and gamma ray fluxes.   
In our first model (model I) we assume a proton population, in the jet rest frame, with spectral indices of p1 = 1.2 and p2 = 2.2 before and after the break energy $E_{brk}$ as follows :

\beq
\frac{dN_p}{dE_p} = \left\{
\begin{array}{lr}
 K ~ E_p^{-p1}, ~~~ E_p < E_{brk} \\
 
 K ~ E_{brk}^{(p2-p1)} ~ E_p^{-p2}, ~~~ E_p \geq E_{brk}
\end{array}
\right.
\eeq
where $K$ is the normalization constant of the proton spectrum. 
We have assumed the knot region is spherical with radius R = 1kpc and the uniform magnetic field (B) within this region is 30mG. 

In case of Bohm diffusion using Eq.(\ref{t_esc}) the escape time of protons is found to be greater than their synchrotron cooling time for the entire range of proton energy taken into consideration for $B=30$mG. 

For both of our models the age of the jet is assumed to be $ 1.4 \times 10^{7}$yr. For $E_p \geq 10^{16}$eV the synchrotron cooling time becomes smaller than the age of the jet (for example at $E_p = 10^{17}$eV, $t_{esc} = 1.3 \times 10^{9}$ yr and $t_{sync} = 1.6 \times 10^{6}$yr), 
which makes the proton spectrum steeper by $E_p^{-1}$. Thus above the break energy the proton spectrum is proportional to $E^{-2.2}$.

\par
%While passing through a region of high magnetic field these shock accelerated protons radiate synchrotron emission. 

The photons emitted in synchrotron emission of the accelerated protons can be visible from earth only if the beamed radiation is along the observer's 
line of site in observer's rest frame. If the radiated photons in the jet frame are at an angle of $\theta_{ob}$ with respect to an 
observer on earth, then the Doppler factor $\delta_D$ of the jet frame is calculated as
\beq
\delta_D=\frac{1} {\Gamma_j(1-\beta_j \cos{\theta_{ob}})}
\eeq
where $\Gamma_j$ and $\beta_j$ are the Lorentz factor and dimensionless velocity of jet rest frame with respect to us.

\par
In our model the jet is moving with a Lorentz factor $\Gamma_j$ = 3.0 and the viewing angle of emitted photons with respect 
to an observer on earth is $\theta_{ob} = 45^{o}$ which give $\delta_D = 1.0$. The synchrotron radiation emitted by protons 
is shown in Fig.\ref{fig1} with black solid line which shows that the proton synchrotron spectrum can explain 
the observed X-ray as well as the gamma ray flux. 

In this model the total energies in the magnetic field and in cosmic ray protons are $W_{j,B} = 4.9 \times 10^{60}$erg and $W_{j,p} = 6.9 \times 10^{57}$erg which give jet powers $L_{j,B} = 1.1 \times 10^{46}$erg/sec and $L_{j,p} = 1.6 \times 10^{43}$erg/sec respectively.
 The maximum proton energy required to explain the observed gamma ray flux is $5 \times 10^{19}$eV. 
 
\par
%In our model I we have assumed proton propagation in Bohm diffusion limit which we are not assuming for our model II.
We propose a second model where the injected proton spectrum has a spectral index of 1.5 (model II). In this model the ambient magnetic 
field is much less than that in model I. In our model II the magnetic field ($B$) is assumed to be $10$mG and the size of the knot region ($R$) is 0.8kpc. Escape time scale calculated using Eq.(\ref{t_esc2}) shows that the loss due to
escape of protons from the emission region is dominant over the loss due to synchrotron cooling. The escape time scale become shorter than the age of the jet 
for proton energy for $E_p \geq 10^{14}$eV (at $10^{14}$eV, $t_{esc} = 1.4 \times 10^{7}$yr and $t_{sync} = 1.4 \times 10^{10}$yr). 
Above this break energy the proton spectrum has a spectral index of 2.0 as $t_{esc}\propto E_p^{-0.5}$ in Eq.(\ref{t_esc2}). The synchrotron flux in model II is shown in our Fig. \ref{fig1} with dashed line and compared with the observed fluxes. The observed X-ray and $\gamma$ ray fluxes are well explained
 by our model II.

The values of the bulk Lorentz factor %$\Gamma_j = 3$ and 
and jet angle %$\theta_{ob} = 45^{o}$ which make 
are assumed to be the same as in our model I which give the Doppler factor $\delta_D = 1.0$.
The jet energy in magnetic and cosmic ray protons are $W_{j,B} = 2.6 \times 10^{59}$erg and $W_{j,p} = 4.6 \times 10^{58}$erg respectively. Therefore the the required jet luminosity in the magnetic field is $L_{j,B} = 5.9 \times 10^{44}$erg/sec and in cosmic ray protons it is $L_{j,p} = 1 \times 10^{44}$erg/sec.
Thus the energies in the magnetic field and in the cosmic ray protons are in $6:1$ ratio. The parameters of both the models are listed in table.\ref{tab1}. 

In both of our models we require a hard proton acceleration spectrum, for model I p1 = 1.2 and for model II p2 = 1.5 before the break energy.   
We note that to explain the the gamma ray emission from Fermi detected gamma ray blazars by proton synchrotron radiation \citet{bot} have 
considered a quite hard injection spectrum of protons. They have taken spectral index of 1.3 for the accelerated proton spectrum to justify the gamma ray emission from PKS 0420$-$01 by synchrotron radiation from relativistic protons. For PKS 1510-089 an accelerated proton spectrum with a spectral index of 1.7 has been used which is consistent with the observed gamma ray flux.

\begin{table}
 \caption{Parameter values for our model I and model II.}
 \label{tab1}
 \begin{tabular}{@{}lccc}
  \hline
  Parameter & Symbol & Our Model I & Our Model II \\
  \hline
  Knot A size(cm) & R & $3.2 \times 10^{21}$ &  $2.5 \times 10^{21}$\\
  Jet Lorentz Factor & $\Gamma_j$ & 3 & 3\\
  Doppler Factor & $\delta_D$ & 1.0 & 1.0\\
  Jet angle & $\theta_{ob}$ & 45$^{o}$ & 45$^{o}$\\
  Magnetic Field(mG) & B & 30 & 10 \\
  %Escape time(yr) & $t_{esc}$ & $1 \times 10^3$ \\
  %Synchrotron Cooling time time(yr) & $t_{sync}$ & $8.7 \times 10^2$ \\
  \hline
  Low Energy Proton Spectral Index & p1 & 1.2 & 1.5 \\
  High Energy Proton Spectral Index & p2 & 2.2 & 2.0 \\
  Minimum Proton Lorentz Factor & $\gamma_{min}$ & $10^5$ & $10^3$ \\
  Maximum Proton Lorentz Factor & $\gamma_{max}$ & $5 \times 10^{10}$ &  $6.3 \times 10^{10}$\\
  Break proton Lorentz Factor & $\gamma_{brk}$ & $1.1 \times 10^{7}$& $10^{5}$\\ 
  \hline
  Jet power in proton (erg/sec) & $L_{j,p}$ & $ 1.6 \times 10^{43}$ & $1.04 \times 10^{44}$ \\
  Jet power in magnetic field (erg/sec) & $L_{j,B}$ & $ 1.1 \times 10^{46}$ & $5.9 \times 10^{44}$ \\
  \hline
 \end{tabular}
\end{table}

\section{Discussion}
The proton synchrotron models discussed in our work show that the X-ray and 
gamma ray fluxes from knot A of 3C 273 may have a common origin in synchrotron emission of the accelerated protons. 
We require a flat proton injection spectrum with spectral index of 1.2 (model I) and 1.5 (model II) before the break energy.
In our model I the protons are assumed to diffuse in Bohm limit while for our
model II an energy dependent escape time has been used. 
In our model I the synchrotron loss dominates over the escape loss and $t_{sync}$ is shorter than the age of the jet above the break energy at $10^{16}$eV resulting in a steeper spectrum of $E_p^{-2.2}$ above this energy.  
In our model II the escape time of the cosmic ray protons is shorter than their synchrotron cooling time. Above the break energy the escape time becomes shorter than the age of the jet as a result the spectrum steepens to $E_p^{-2}$ from $E_p^{-1.5}$.

\par
 
%According to \citet{ebi} protons can be accelerated to $10^{21}$eV by wakefield and pondermotive force in AGN jets. 
%The maximum possible energy achieved by particles in wakefield acceleration is calculated as follows :
%\beq
%E_{cut} = 2.9 \times 10^{22} ~ Z ~ \Bigg({\frac{\dot{m}}{0.1}}\Bigg)^{4/3}  \Bigg({\frac{m}{10^8}}\Bigg)^{2/3} ~ eV
%\eeq

%where $m$ is the mass of the black hole in units of solar mass and $\dot{m}$ is the accretion rate normalized to the critical accretion rate.  
%$Z$ is the charge of the particle which is 1 for proton. For 3C 273 $m = 10^{9}$ and the value of $\dot{m}$ is between  $0.4$ and $0.13$ \citep{malkan} which gives the maximum energy of the protons $\sim 10^{22}$ eV in the jet frame. 

The maximum energies of the cosmic ray protons required in our models are $5 \times 10^{19}$eV and $6.3 \times 10^{19}$eV respectively. %In general the knots and hotspots  
%of large scale AGN jets are speculated to be acceleration sites extremely high energy protons where protons can be accelerated up to $10^{20}$eV. 
If cosmic ray protons are accelerated in the jets of AGN \citep{murase,ebi} 
upto $\sim 10^{20}$ eV their synchrotron radiation could be the origin of the X-ray and gamma ray emissions observed from knot A of 3C 273.   
An important aspect of the proton synchrotron model is the acceleration site of the protons does not have to be located within the emission site of the synchrotron photons. As the accelerated protons lose energy at a much slower rate compared to the electrons they can travel long distances from their acceleration site/sites before losing energy significantly.  
%The proton energy 
%in the comoving jet frame is $E_{max} = E_{max}' \times (1+z)/\Gamma$. The redshift of 3C 273 is $z = 0.158$. 
%For $\dot{m} = 0.4$ the maximum energy up to which protons can be accelerated in the jet frame is $E_{max} = 2.2 \times 10^{22}$eV. In case of $\dot{m} = 0.13$ $E_{max} = 4.7 \times 10^{21}$eV. 
%We note that in case of 3C 273 the maximum attainable energy of protons in wakefield acceleration mechanism is about three order of magnitude higher than that the maximum energy required for our models.
%Moreover the pondermotive force can accelerate charged particles over a long distance in AGN jets. If wakefield acceleration mechanism is the underlying 
%process through which particles are accelerated in the large scale jet of 3C 273 over a long distance, then these accelerated particles can possibly 
%be the sources of the ultrahigh energy protons in the knot A which emit synchrotron radiation in X-ray to gamma ray frequency. 
%In this acceleration mechanism the particles acquire a power law spectrum with spectral index of 2. But in case wakefield has two or three dimensional 
%feature the spectrum of accelerated particles become harder than 2.

\par
In our model I the total luminosity required to explain the observed X-ray and gamma ray fluxes by the synchrotron radiation of protons is $10^{46}$erg/sec which is about 10\% of the Eddington's luminosity of 3C 273. 
For model II the total jet luminosity required is $\sim 10^{45}$erg/sec. In our model I the jet power in magnetic field is about three orders of
magnitude higher than that in cosmic ray protons and this factor reduces to 
6 in our model II.

%\end{verbatim}

\end{document}